\documentclass[twocolumn,aps,prl,showpacs,floatfix]{revtex4}

\usepackage{graphicx}

\begin{document}

\title{Density of states in random lattices with translational invariance}

\author{Y. M. Beltukov}
\author{D. A. Parshin}
\affiliation{Saint Petersburg State Polytechnical University, 195251 Saint Petersburg, Russia}

\date{\today}

\begin{abstract}
We propose a random matrix approach to describe vibrations in disordered systems. The dynamical matrix $M$ is  taken in the form $M=AA^T$ where $A$ is a real random matrix. It guaranties that $M$ is a positive definite matrix. This is necessary for mechanical stability of the system. We built
matrix $A$ on a simple cubic lattice with translational invariance and interaction between nearest neighbors. It was found that for a certain type of disorder acoustical phonons cannot propagate through the lattice and the density of states $g(\omega)$ is not zero at $\omega=0$. The reason is a breakdown of affine assumptions  and inapplicability of the macroscopic elasticity theory. Young modulus goes to zero in the thermodynamic limit. It reminds of some properties of a granular matter at the jamming transition point. Most of the vibrations are delocalized and similar to diffusons introduced by Allen, Feldman et al., Phil. Mag. B {\bf 79}, 1715 (1999). We show how one can gradually return rigidity and phonons back to the system increasing the width of the so-called {\em phonon gap} (the region where $g(\omega)\propto\omega^2$). Above the gap the reduced density of states $g(\omega)/\omega^2$ shows a well-defined Boson peak which is a typical feature of glasses. Phonons cease to exist above the Boson peak and diffusons are dominating. It is in excellent agreement with recent theoretical and experimental data.
\end{abstract}

\maketitle

In a sharp contrast to electronic properties the nature of vibrations in disordered systems is still poorly understood though these excitations are responsible for such important phenomena  as specific heat, thermal conductivity, propagation of sound and elastic properties. Solid amorphous dielectrics (glasses) are the most common example among these systems. Others are polymers, liquids, granular media, etc.

Low frequency plane long wave acoustical phonons (or Goldstone modes)
are the most extensively studied delocalized excitations
in many disordered materials. They propagate through the media ballistically with a speed of sound. But we yet do not know what is the upper frequency where one can still use this useful notion. One can find in the literature hot debates about existence of high frequency phonons in glasses.

Some time ago a new type of delocalized vibrations in disordered media was introduced that was called {\em diffusons}~\cite{Nature}. These are vibrations extending through the system by means of diffusion. It is an important class of excitations which  occupy in glasses the dominant part of the spectrum. The diffusons may be responsible for the thermal conductivity of glasses above the plateau. According to~\cite{Pohl} the heat in glasses above 20\,K is transmitted by means of a random walk of vibrations from one atom to its nearest neighbors.

In the last years it was discovered that amorphous materials as diverse as granular media, foams, emulsions, and colloidal suspensions can jam into a rigid, disordered state where they withstand finite stresses before yielding~\cite{Liu,jam}. At the point of jamming transition these systems are marginally stable but there is a breakdown of affine assumptions that underlies their rich mechanics near jamming. Far away from jamming the density of states (DOS) shows Debye-like behavior $g(\omega)\propto\omega^{d-1}$. But as the jamming point is approached, both the structure of the modes and the DOS exhibit surprising features. In particular $g(\omega)$ at low frequencies is strongly enhanced and becomes essentially flat at $\omega=0$~\cite{jamming dos}. One can conclude that low frequency phonons disappear at this point and the macroscopic elasticity theory becomes inapplicable. It was found that the low frequency modes are neither plane-wave-like nor localized~\cite{jam3d} and they are similar to diffusons~\cite{Vit}.

Recently using a random matrix theory we developed a model~\cite{our1} which we believe can describe typical properties of vibrations in disordered systems. Here we present an example of a $3d$ disordered lattice with translational invariance where $g(\omega)$  is  nonzero at $\omega\to 0$ and almost all vibrations are delocalized. We show that affine assumptions are violated and no phonons exist in the lattice. These properties are similar to those of disordered systems at the jamming transition point. However there is important difference between these two systems. We show how one can continuously return rigidity and phonons back to the system increasing the width of the so-called {\em phonon gap} (the region where $g(\omega)\propto\omega^2$).

The  vibrational frequencies squared $\omega_i^2$ for mechanical system of $N$ particles are the eigenvalues of the dynamical matrix $M_{ij}=\Phi_{ij}/\sqrt{m_im_j}$, where $\Phi_{ij}$ is a force constant matrix and $m_i$ are the particle masses. The matrices $M$ and $\Phi$ are real, symmetric and {\em positive definite} matrices $N\times N$ (scalar model). The last condition is important. It ensures mechanical stability of the system.

To describe vibrations in disordered systems we want to use the methods of random matrix theory.
One can always present every real symmetric and positive definite matrix $M$ in the form~\cite{PositiveDefine,Chalker}
\begin{equation}
     M = AA^T.
\end{equation}
Here $A$ is some real matrix of a general form. And vice versa for every real matrix $A$
the product $AA^T$ is always a positive definite symmetric matrix. In this paper we are going to consider disordered systems where $\omega^2$ are eigenvalues of matrix $AA^T$, with $A$ taken randomly.

Distribution of eigenvalues of matrix $AA^T$  was found in~\cite{Marchenko}. The authors investigated the case where $A$ is a real random matrix with independent elements with zero mean $\left\langle A_{ij}\right\rangle = 0$ and equal dispersions $\left\langle A_{ij}^2\right\rangle = V^2$ (Wishart ensemble~\cite{Wishart}).
For a squarte matrix $A$ ($N\times N$) the distribution of frequencies $g(\omega)$
for $N\gg 1$ has a {\em quarter-circle} form (with radius $\propto\sqrt{N}$). As a result $g(\omega)$ is a {\em constant} at $\omega=0$. Matrices $AA^T$ were investigated in the theory of financial markets~\cite{Plerov}, complex networks~\cite{Barthelemy} and wireless communications~\cite{Tulino}. For vibrations in random-field spin chains (without translational invariance) this approach was used in~\cite{Chalker}.

In Wishart ensemble each matrix element $M_{ij}$ is in general not zero
\begin{equation}
M_{ij}=\sum_k A_{ik} A_{jk}.
\label{vst5rf}
\end{equation}
It corresponds to the infinite-range interaction between particles.
A similar result for the density of states was found for {\em sparse} random matrices $A$ with only $n\ll N$ non-zero random elements in each row. In the limit $N\gg n\gg 1$, $g(\omega)$ also approaches the quarter-circle form~\cite{our1}
\begin{equation}
     g(\omega) = \frac{N}{\pi nV^2}\sqrt{4nV^2-\omega^2}, \quad 0 < \omega <  2V\sqrt{n}
     \label{duvb6}
\end{equation}
but with radius independent of the number of particles $N$.

Another possibility is to build random matrix $A$ on a lattice.
Let us consider a simple cubic lattice with $N$ particles.
Each particle has its unique integer index $i$ which takes values from $1$ to $N$. We build the matrix $A$ as follows. The element $A_{ij}$  will have a nonzero random value if $i$th and $j$th particles are the nearest neighbors ($i\ne j$) or  it is the same particle ($i=j$). All other elements $A_{ij}=0$.

\begin{figure}[!h]
     \includegraphics[totalheight=4.5cm,keepaspectratio]{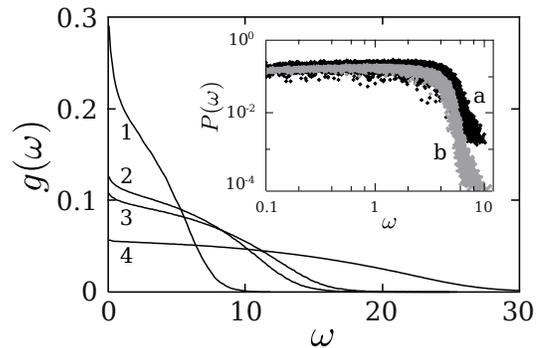}
     \caption{The normalized DOS $g(\omega)$ in a simple cubic lattice. (1) scalar nnm, (2) vector nnm, (3) scalar nnnm, (4) vector nnnm (nnm --- nearest neighbor model, nnnm --- next nearest neighbor model (up to the third shell)). Inset:  participation ratio $P(\omega)$ for a) $N=10^3$ and b) $N=27^{3}$.}
     \label{fig:dosprmu0}
\end{figure}

In~\cite{our1} it was  considered the case of so-called  {\em pinned} latices. Diagonal and non-diagonal elements of matrix $A$ were independent random numbers without any correlation. Thus each particle was randomly connected not only with its neighbors but also with {\em space}.  However the Goldstone modes (phonons) cannot propagate through such lattice. For existence of phonons it is necessary  to satisfy also conditions of {\em translation invariance}
\begin{equation}
\sum_iM_{ij}=\sum_jM_{ij}=0
\label{s6cv}
\end{equation}
(we consider $m_i=\mbox{const}$).
It ensures that the Newton equations for particle displacements $u_i$ have solutions $u_i=\mbox{const}$ for $\omega=0$ what is the necessary condition for existence of Goldstone modes.

In this paper we consider a simple cubic lattice with translational invariance. We take non-diagonal elements of matrix $A$ for nearest neighbors as independent random numbers, but diagonal elements $A_{ii}$ will satisfy  conditions
\begin{equation}
A_{ii}= -\sum\limits_{j\ne i} A_{ji} .
\label{67vb}
\end{equation}
Then according to Eq.(\ref{vst5rf}) the Eq.~(\ref{s6cv}) will be also satisfied.

Fig.\ref{fig:dosprmu0} shows the normalized DOS $g(\omega)$ of cubic lattice $20\times 20\times 20$ with translational invariance (we used periodic boundary conditions). The average values of non-diagonal elements $\left<A_{ij}\right>=0$ and dispersion $\left<A_{ij}^2\right>=1$ (for Gauss distribution). The diagonal elements $A_{ii}$ were calculated using Eq.~(\ref{67vb}). The results are surprising. We do not see the expected phonon modes with their $g_{ph}(\omega)\propto \omega^2$ for $\omega\to 0$. On the contrary for scalar model with nearest neighbor interactions $g(\omega)$ even increases at small $\omega$. This result is almost {\em identical} to $g(\omega)$ of similar pinned lattice~\cite{our1}.

The increase of $g(\omega)$ at small $\omega$ is due to weak logarithmic singularity superimposed on a smooth dependence $g_{\rm sm}(\omega)$. It exists also for a sparse random matrix $A$ with sufficiently small coordination number $n$~\cite{our1}. However with increasing $n$ the singularity is suppressed  and $g(\omega)$ approaches the quarter-circle form, Eq.~(\ref{duvb6}), i.e. becomes constant at small $\omega$~\cite{our1}. We verified that flatness of $g(\omega)$ at small $\omega$ is also the case in our cubic lattice  if we increase the number of interacting neighbors or switch from scalar to vector model or both (see Fig.\ref{fig:dosprmu0}). Since this singularity is not important for the following consideration we will consider below the scalar model with nearest neighbor interaction.

To understand what kind of vibrations we have in our lattice we show on the Fig.\ref{fig:dosprmu0}  the participation ratios calculated  according to
\begin{equation}
\label{cvr9}
     P(\omega)=\left[N\sum\limits_{i=1}^{N}e_i^4(\omega)\right]^{-1}.
\end{equation}
Here $e_i(\omega)$ is $i$th coordinate of the normalized eigenvector with frequency $\omega$. As can be seen all modes with exception of high frequencies are {\em delocalized}. They have  $P(\omega)\approx 0.2$ which is independent of the system size. Similar results were obtained for pinned lattices~\cite{our1}. In both cases we identified these excitations as diffusons. It was shown that the energy transfer through the lattice has character of diffusion
and the level spacing distribution obeys Wigner-Dyson statistics. It also indicates  mode delocalization.
It is remarkable that  diffusons turned out to be not sensible to the presence or absence of translational invariance. These results will be published elsewhere.

\begin{figure}[!h]
     \includegraphics[totalheight=5cm,keepaspectratio]{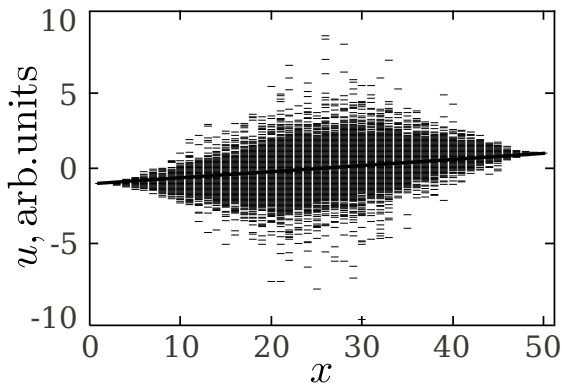}
     \caption{Non-affine particle displacements in different atomic layers. Solid line is the prediction of the elasticity theory.}
     \label{fig:ss88}
\end{figure}
Absence of phonons signals that macroscopic elasticity theory  becomes inapplicable. It is similar to systems at the jamming transition point.
To check this idea we stretched our sample $50\times 50\times 50$ for two opposite ends with {\em displacements} $u=\pm 1$ at the ends correspondingly. Then we calculated the particle displacements $u$ in each atomic layer (2500 dashes). They are shown on Fig.\ref{fig:ss88}. The displacements
do not obey the elasticity theory predictions shown by the solid line.
The fluctuations in displacements of most of the particles are of the order of  unity.
It means that affine assumptions are violated and hence no low frequency phonons (plane waves) can exist in
these lattices in spite of the translational invariance.
Non affine displacements and  violation of elasticity theory were also found in computer simulations of small amorphous bodies formed by weakly polydisperse Lennard-Jones beads~\cite{Tanguy}.

\begin{figure}[htb]
     \includegraphics[totalheight=4.5cm,keepaspectratio]{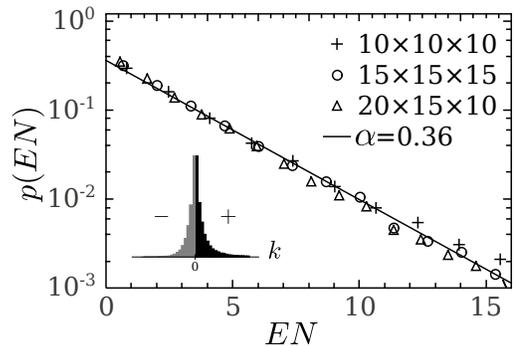}
     \caption{The distribution of Young modulus $E$ (multiplied by number of particles $N$) for three different samples. Inset: the distribution of elastic constants, $\left<k\right>=0.5$,
     $\left<k^2\right>=5.5$.}
     \label{fig:E}
\end{figure}
To measure the Young modulus of the lattice we stretched our sample at two opposite ends with {\em forces} $f=\pm 1$ per each particle (at the end) correspondingly. The mean sample extension was calculated as difference between mean displacements of particles at the ends
\begin{equation}
     \Delta L=\langle u_i\rangle_{f=1}-\langle u_i\rangle_{f=-1}.
\end{equation}
Then the Young modulus (in our scalar model) is
\begin{equation}
     E=\frac{|f|L}{a_0^2|\Delta L|},
\end{equation}
where $L$ is the sample length and $a_0=1$ is the lattice constant. First we found that the Young modulus is a strongly fluctuating quantity. It depends on the choice of the random matrix $A$. Secondly, the average value $\left<E\right>\propto 1/N$ and decreases with the system size. In the thermodynamic limit ($N\to\infty$) the average Young modulus is zero. It is similar to results of~\cite{epitome} for granular matter where static shear and bulk moduli of the disordered system of interacting particles approach zero at the jamming transition point.
Fig.~\ref{fig:E} shows the distribution of Young modulus $E$ multiplied by the number of particles $N$ for 3 different samples. All points  lay perfectly  on the straight line (in semi-logarithmic scale). It means that distribution function $p(E)$ has a form
\begin{equation}
p(E)=\alpha N e^{-\alpha EN}, \quad \alpha\approx 0.36 \approx 1/e.
\end{equation}
The distribution of elastic constants $k=-M_{i\ne j}$ is shown on the same figure ($N=50^3$).
There are a lot of negative springs (with $k<0$) in the system (about 45\%).

The similarity of our disordered lattice with granular media can be also found in distribution of random forces in deformed lattice. We have loaded our cubic sample $20\times 20\times 20$ with gravitational forces  equal for each particle.  Then at the bottom of the sample the distribution of  contact forces was found to be exponential, similar to what was measured in a granular matter~\cite{force}.

\begin{figure}[htb]
     \includegraphics[totalheight=4.5cm,keepaspectratio]{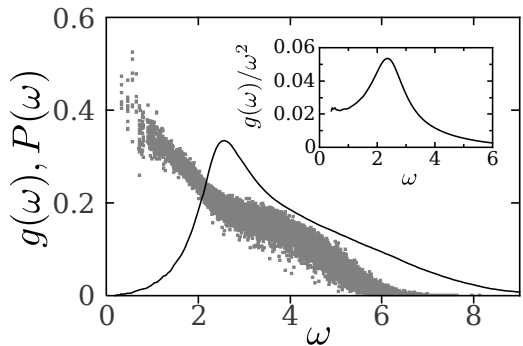}
     \caption{The normalized DOS $g(\omega)$ (solid line) and participation ratio $P(\omega)$ (dots) for dynamical matrix $M=AA^T+\mu M_0$ with $\mu=1$. Inset shows the Boson peak in $g(\omega)/\omega^2$.}
     \label{fig:dosprmu1}
\end{figure}
To get phonons we should introduce  a finite rigidity into the system. For that let us consider a dynamical matrix of the form
\begin{equation}
M=AA^T+\mu M_0.
\end{equation}
Here matrix $A$ is the same random matrix in a $3d$ cubic lattice with translational invariance. Matrix $M_0$ is a regular dynamical matrix for the same lattice with unit masses and all spring constants (between the nearest neighbors) equal to unity.

The results are shown on Fig.\ref{fig:dosprmu1} for $\mu=1$ and $N=20^3$. Now  $g(\omega)\propto\omega^2$ at $\omega\to 0$. So we revive phonons at small $\omega$ below the maximum.
The participation ratio $P(\omega)$ ($N=27^3$) in this frequency range has increased by more than factor of two (up to the crystal value) what is a clear signature of existence of well defined Goldstone modes. At higher frequencies (above the maximum) we have the same delocalized excitations as on Fig.\ref{fig:dosprmu0} i.e. diffusons. On the inset we can also see that in the transition region between phonons and diffusons the reduced density of states $g(\omega)/\omega^2$ shows a well defined peak. This is a well known Boson peak which is a typical feature of glasses~\cite{boson1}. We see that in our model phonons cease to exist above the Boson peak. This is in agreement with results of papers~\cite{Laermans,Ruffle,Schober} derived for glasses. Boson peak in our disordered lattice  has a non-phonon origin. We do not believe it is related to Van Hove singularity shifted due to disorder as was found in~\cite{Schirmacher,Taraskin}.

\begin{figure}[!h]
     \includegraphics[totalheight=5cm,keepaspectratio]{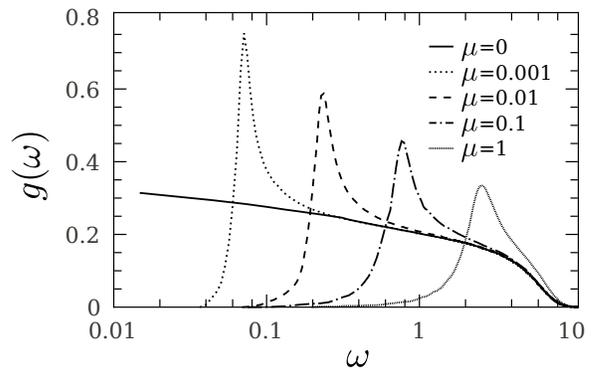}
     \caption{The normalized DOS $g(\omega)$ corresponding to dynamical matrix $M=AA^T+\mu M_0$ with different $\mu$.}
     \label{fig:gwmu}
\end{figure}

As a result introducing finite values of $\mu$ we open in the system a {\em phonon gap} --- Fig.~\ref{fig:gwmu}.
It is a frequency range where $g(\omega)\propto\omega^2$ and phonons exist as well defined excitations. Above the gap phonons cease to exist and diffusons are the only delocalized excitations.
The width of the gap and the frequency of the Boson peak increases $\propto\sqrt{\mu}$. Similar changes of the
spectrum take place when compressing the marginally jammed solid to higher packing fractions~\cite{jamming dos}.

Concluding we have shown that for some class of disorder in  lattices with translational invariance acoustical phonons cannot propagate through the system since the macroscopic elasticity theory and affine assumptions become inapplicable. The major part of vibrations are delocalized diffusons. Their DOS is not zero at $\omega=0$. From this point our disordered lattices have similar properties with granular systems at the jamming transition point.
We have shown that this unusual behavior is due to existence of high concentration of {\em negative springs} which makes our lattice extremely soft. It is different from granular jammed systems where a coordination number (the mean number of contacts per particle) plays an important role.

Introducing a finite rigidity we return phonons back to the system increasing the width of the phonon gap. Inside the gap phonons are well defined excitations. Outside the gap the diffusons are dominating and phonons cease to exist. In the transition region between phonons and diffusons our system shows a well-defined Boson peak which is a typical feature for glasses. As a result our simple scalar random matrix model is able to reproduce typical properties of various disordered systems starting from granular matter at the jamming transition point, to jammed systems and finally to real glasses.

We are very grateful to V.L.~Gurevich and V.I.~Kozub for many stimulating discussions and critical reading of the manuscript.



\begin{thebibliography}{99}

\bibitem{Nature} P.\,B. Allen, J.\,L. Feldman, J. Fabian, F. Wooten. Phil.~Mag. B {\bf 79}, 1715 (1999).

\bibitem{Pohl}  D.\,G. Cahill, S.\,K. Watson, R.\,O. Pohl, Phys.~Rev. B {\bf 46}, 6131 (1992).

\bibitem{Liu} A.\,J. Liu, and S.\,R. Nagel, Nature {\bf 396}, 21 (1998).

\bibitem{jam} M. van~Hecke, J.~Phys.:~Condens.~Matter, {\bf 22}, 033101 (2010).

\bibitem{jamming dos} L.\,E. Silbert, A.\,J. Liu, and S.\,R. Nagel  Phys.~Rev.~Lett. {\bf 95}, 098301 (2005).

\bibitem{jam3d} L.\,E. Silbert, A.\,J. Liu, and S.\,R. Nagel, Phys.~Rev. E {\bf 79}, 021308 (2009).

\bibitem{Vit} V. Vitelli, N. Xu, M. Wyart, A.\,J. Liu, S.\,R. Nagel, Phys.~Rev. E {\bf 81}, 021301 (2010).

\bibitem{our1} Y.\,M. Beltukov and D.\,A. Parshin, Physics of the Solid State {\bf 53}, 151 (2011) (Fizika Tverdogo Tela, {\bf 53}, 142 (2011)).

\bibitem{PositiveDefine} R. Bhatia. Positive Definite Matrices. Princeton University Press, Princeton (2007). 264~ñ.

\bibitem{Chalker} V. Gurarie, and J.\,T. Chalker, Phys.~Rev.~Lett. {\bf 89}, 136801 (2002);
Phys.~Rev. B {\bf 68}, 134207 (2003).

\bibitem{Marchenko} V.\,A. Mar\u{c}enko and L.\,A. Pastur, Math. USSR-Sbornik, {\bf 1}(4), 457 (1967).

\bibitem{Plerov} V.~Plerou, P.~Gopikrishnan, B.~Rosenow, L.\,A.\,N. Amaral, T. Guhr, H. Stanley.
          Phys.~Rev.~E {\bf 65}, 066126 (2002).

\bibitem{Barthelemy} M. Barthelemy, B. Gondran, E. Guichard. Phys.~Rev. E {\bf 66}, 056110 (2002).

\bibitem{Tulino} A.\,M. Tulino, S. Verd\`u. Random Matrix Theory and Wireless Communications. Fundations and
           Trends in Communications and Information Theory. {\bf 1}, 1 (2004).

\bibitem{Wishart} J. Wishart. Biometrika, {\bf 20 A}, 32 (1928).

\bibitem{Tanguy} A. Tanguy, J.\,P. Wittmer, F. Leonforte, and J.\,-L. Barrat, Phys.~Rev. B {\bf 66}, 174205 (2002).

\bibitem{epitome} C.\,S. O'Hern, L.\,E. Silbert, A.\,J. Liu, S.\,R. Nagel Phys.~Rev. E {\bf 68}, 011306 (2003).

\bibitem{force} D.\,M. Mueth, H.\,M. Jaeger, and S.\,R. Nagel, Phys.~Rev. E {\bf 57}, 3164 (1998).

\bibitem{boson1} V.\,L. Gurevich, D.\,A. Parshin, H.\,R. Schober, Phys.~Rev. B {\bf 67}, 094203 (2003);
D.\,A. Parshin, H.\,R. Schober, V.\,L. Gurevich, Phys.~Rev. B {\bf 76}, 064206 (2007).

\bibitem{Laermans} D.\,A. Parshin, C. Laermans, Phys.~Rev. B {\bf 63}, 132203 (2001).

\bibitem{Ruffle} B. Ruffle, D.\,A. Parshin, E. Courtens, and R. Vacher, Phys.~Rev.~Lett. {\bf 100}, 015501 (2008).

\bibitem{Schober} H.\,R. Schober, J. Phys.: Condens. Matter {\bf 16}, S2659 (2004).

\bibitem{Schirmacher} W. Schirmacher, G. Diezemann, and C. Ganter, Phys.~Rev.~Lett. {\bf 81}, 136 (1998).

\bibitem{Taraskin} S.\,N. Taraskin, Y.\,L. Loh, G. Natarajan, and S.\,R. Elliott, Phys.~Rev.~Lett. {\bf 86},
1255 (2001).


\end{thebibliography}
\end{document}